\documentclass{PoS}

\usepackage{cite}
\usepackage{axodraw}
\usepackage{epsfig}

\newcommand{\gsim}
{\mathrel{\raisebox{-.3em}{$\stackrel{\displaystyle >}{\sim}$}}}
\def\asymp#1%
{\mathrel{\raisebox{-.4em}{$\widetilde{\scriptstyle #1}$}}}

\def\Nequal#1%
{\mathrel{\raisebox{-.5em}{$\stackrel{=}{\scriptstyle\rm#1}$}}}
\newcommand{\dsl}[1]{\not \hspace{-0.7mm}#1}
\def\dsl{\mathpalette\make@slash}
\def\make@slash#1#2{\setbox\z@\hbox{$#1#2$}%
  \hbox to 0pt{\hss$#1/$\hss\kern-\wd0}\box0}

\def\beq{\begin{equation}}
\def\eeq{\end{equation}}
\def\bit{\begin{itemize}}
\def\eit{\end{itemize}}
\def\beqar{\begin{eqnarray}}
\def\eeqar{\end{eqnarray}}
\def\barr#1{\begin{array}{#1}}
\def\earr{\end{array}}
\def\bfi{\begin{figure}}
\def\efi{\end{figure}}
\def\btab{\begin{table}}
\def\etab{\end{table}}
\def\bce{\begin{center}}
\def\ece{\end{center}}

\def\text{\textstyle}

\newcommand{\eqintext}{}

\def\reffi#1{\mbox{Figure~\ref{#1}}}

\def\citere#1{\mbox{Ref.~\cite{#1}}}
\def\citeres#1{\mbox{Refs.~\cite{#1}}}

\newcommand{\TeV}{\unskip\,\mathrm{TeV}}
\newcommand{\GeV}{\unskip\,\mathrm{GeV}}

\newcommand{\ri}{{\mathrm{i}}}

\newcommand{\rT}{{\mathrm{T}}}

\def\mathswitchr#1{\relax\ifmmode{\mathrm{#1}}\else$\mathrm{#1}$\fi}

\newcommand{\PW}{\mathswitchr W}

\newcommand{\PH}{\mathswitchr H}
\newcommand{\Pe}{\mathswitchr e}
\newcommand{\Pne}{\mathswitch \nu_{\mathrm{e}}}

\newcommand{\Pnmubar}{\mathswitch \bar\nu_{\mu}}

\newcommand{\Pb}{\mathswitchr b}
\newcommand{\Pbbar}{\mathswitchr{\bar b}}
\newcommand{\Pp}{\mathswitchr p}
\newcommand{\Pt}{\mathswitchr t}
\newcommand{\Ptbar}{\mathswitchr{\bar t}}
\newcommand{\Pep}{\mathswitchr {e^+}}

\newcommand{\Pmum}{\mathswitchr {\mu^-}}

\newcommand{\Pl}{l}

\def\mathswitch#1{\relax\ifmmode#1\else$#1$\fi}

\newcommand{\Mt}{\mathswitch {m_\Pt}}

\newcommand{\GW}{\Gamma_{\PW}}

\newcommand{\Gt}{\Gamma_{\Pt}}

\def\solid{\raise.9mm\hbox{\protect\rule{1.1cm}{.2mm}}}
\def\dash{\raise.9mm\hbox{\protect\rule{2mm}{.2mm}}\hspace*{1mm}}

\def\ie{i.e.\ }

\newcommand{\pt}{\mathswitch {p_{\mathrm{T}}}}

\newcommand{\ttb}{{\Pt\bar\Pt}}
\newcommand{\wwbb}{{\PW\PW\Pb\bar\Pb}}

\newcommand{\wwfs}{\PW^+\PW^-\Pb\bar\Pb}
\newcommand{\leptfs}{{\Pne\Pep\Pmum\Pnmubar\Pb\bar\Pb}}

\title{NLO QCD corrections to off-shell \boldmath{$\mathrm{t\bar t}$} production at hadron colliders}

\ShortTitle{NLO QCD corrections to off-shell ${t\bar t}$ production at hadron colliders}

\author{Ansgar Denner\\ 
        Universit\"at W\"urzburg, Institut f\"ur Theoretische Physik und Astrophysik,
        D-97074 W\"urzburg, Germany\\
        E-mail: \email{denner@physik.uni-wuerzburg.de}}

\author{\speaker{Stefan Dittmaier}\\
        Albert-Ludwigs-Universit\"at Freiburg, Physikalisches Institut,
        D-79104 Freiburg, Germany\\
        E-mail: \email{stefan.dittmaier@physik.uni-freiburg.de}}

\author{Stefan Kallweit\\
        Institut f\"ur Theoretische Physik, Universit\"at Z\"urich,
        CH-8057 Z\"urich, Switzerland\\
        E-mail: \email{kallweit@physik.uzh.ch}}

\author{Stefano Pozzorini\\
        Institut f\"ur Theoretische Physik, Universit\"at Z\"urich,
        CH-8057 Z\"urich, Switzerland\\
        E-mail: \email{pozzorin@physik.uzh.ch}}

\abstract{The production of top--antitop-quark pairs 
at hadron colliders is interesting both in its own right as signal process,
but also as background to many searches for new physics.
The corresponding predictions aim at the precision level of few per cent, rendering
not only the inclusion of radiative corrections of the strong and electroweak 
interactions relevant, but also of off-shell and finite-width effects originating
from the top-quark decays $\Pt\to\Pb\PW\to\Pb+\Pl^+\nu_\Pl/q\bar q'$.
We report on a calculation for the full process 
$\Pp\Pp\to\PW^+\PW^-\Pb\bar\Pb\to\leptfs$
at next-to-leading order QCD and discuss the effects of
the finite widths of the top quarks and of the W~bosons
for selected observables. Generically it turns out that
finite-top-width effects are at the per-cent level whenever
the top-quark resonances dominate, but those effects can reach tens
of per cent in off-shell tails. 
Finite-W-width effects, on the other hand, are suppressed to less than
$0.5\%$ whenever the top quarks can become resonant and only become sizeable
in exceptional cases. One such case, however, is the invariant mass
of a bottom quark and the corresponding charged lepton, which result from the same
top-quark decay---an observable that is relevant for precision measurements
of the top-quark mass.}

\FullConference{Loops and Legs in Quantum Field Theory - 11th DESY Workshop on Elementary Particle Physics,\\
		April 15-20, 2012\\
		Wernigerode, Germany}

\begin{document}

\section{Introduction}

After about 17 years of its discovery, experimental studies of the top quark
are still highly interesting, since the top quark is expected to
serve as a key for understanding the fermionic mass hierarchy
and to be potentially sensitive to physics
beyond the Standard Model.  
Evidently, accurate measurements of top--antitop pairs carried out
at the Tevatron and the LHC must be accompanied by
precise calculations.
A survey of LHC results on top-quark physics and corresponding theoretical
predictions can be found in the recent review~\cite{Schilling:2012dx}.
In this article we can only concentrate on the specific issue of
off-shell and finite-width effects and the corresponding literature.

Since many years, {$\Pt\bar\Pt$} production at hadron colliders 
is fully known at next-to-leading-order (NLO) within QCD and electroweak theory.
Beyond fixed order, the resummation of logarithmically enhanced QCD corrections 
got more and more refined as well. 
Most recently, a major step was made towards a full prediction at
next-to-next-to-leading-order QCD upon completing the quark--antiquark 
channel~\cite{Baernreuther:2012ws}, which is dominating $\Pt\Ptbar$
production at the Tevatron.
Most predictions, however, are based on the approximation of
stable (on-shell) top quarks, \ie the top-quark decays, which
proceed into pairs of W~bosons and bottom quarks in the Standard Model,
are ignored.  A first important step towards a full NLO description of
top-pair production and decay was made
in~\citeres{Bernreuther:2004jv,Melnikov:2009dn, Bernreuther:2010ny},
where top-quark decays were treated in a spin-correlated narrow-width
approximation, \ie the top quarks are still on shell, but spin
correlations between production and decay are taken into account.
Recently, first results at NLO QCD for the complete process of
$\PW^+\PW^-\Pb\bar\Pb$ production, with intermediate off-shell top
quarks and including leptonic W-boson decays have been obtained by two
independent groups~\cite{Denner:2010jp,Bevilacqua:2010qb,Denner:2012yc}.

As long as resonant $\ttb$ production dominates an observable,
finite-width effects of the top quark are suppressed to the order
of ${\cal O}(\Gt/\Mt)\sim1\%$. The finite-W-width effects 
even turn out to be suppressed to the level
${\cal O}(\Gt/\Mt\times\GW/m_{\PW})$ in inclusive 
observables~\cite{Denner:2012yc}.
However, in more exclusive measurements, such as
precision $\Mt$ determinations, $\ttb$
backgrounds to new physics that are suppressed by vetoing top
resonances, or the $\ttb$ background to \mbox{$\PH\to ll\nu\nu$}
signals in presence of b-jet vetoes,
the investigation of finite-top-width effects is even more important
since their magnitude is not known a priori.
A first systematic study of finite-top-width effects in exclusive 
observables~\cite{Denner:lh2011}, based on a
comparison of our calculation against the narrow-top-width approximation of
\citere{Melnikov:2009dn}, indicates that finite-top-width corrections to
phenomenologically important observables can range from a few per mille
to tens of per cent.
This also raises the issue of possible non-negligible
effects resulting from the finite width of intermediate W bosons, which
is addressed in detail in \citere{Denner:2012yc}.

In this article we briefly discuss results from the off-shell
calculation~\cite{Denner:2010jp,Denner:2012yc} and compare them
to the narrow-width approximation of \citere{Melnikov:2009dn}, 
in order to work out the genuine finite-width effects of the top quark
and the W~boson.

\section{Features of off-shell \boldmath{$\mathrm{t\bar t}$} production}

\begin{figure}
\unitlength 0.83pt
\SetScale{0.83}
\begin{center}
\begin{picture}(170,135)(-30,0)
\Text (-5.,0)[r]{$\mathrm{\scriptstyle g}$}
\Text (-5.,75)[r]{$\mathrm{\scriptstyle g}$}
\Text (110,0)[l]{$\mathrm{\scriptstyle \bar{b}}$}
\Text (110,15)[l]{$\mathrm{\scriptstyle \mu^-}$}
\Text (110,30)[l]{$\mathrm{\scriptstyle \bar{\nu}_\mu}$}
\Text (110,75)[l]{$\mathrm{\scriptstyle b}$}
\Text (110,45)[l]{$\mathrm{\scriptstyle \nu_e}$}
\Text (110,60)[l]{$\mathrm{\scriptstyle e^+}$}
\Vertex(79,52.5){2.0}
\ArrowLine(79,52.5)(105,45)
\ArrowLine(105,60)(79,52.5)
\Vertex(53,60){2.0}
\ArrowLine(53,60)(105,75)
\Photon(79,52.5)(53,60){2}{3}
\Text (75.5,53.)[rt]{$\mathrm{\scriptstyle W^+}$}
\Vertex(79,22.5){2.0}
\ArrowLine(79,22.5)(105,15)
\ArrowLine(105,30)(79,22.5)
\Vertex(53,15){2.0}
\ArrowLine(105,0)(53,15)
\Photon(79,22.5)(53,15){2}{3}
\Text (75.5,22.)[rb]{$\mathrm{\scriptstyle W^-}$}
\Vertex(26,15){2.0}
\Gluon(0,0)(26,15){3}{4}
\ArrowLine(53,15)(26,15)
\Text (39.5,11)[ct]{$\mathrm{\scriptstyle \bar{t}}$}
\Vertex(26,60){2.0}
\ArrowLine(26,15)(26,60)
\Text (22,37.5)[r]{$\mathrm{\scriptstyle t}$}
\Gluon(0,75)(26,60){3}{4}
\ArrowLine(26,60)(53,60)
\Text (39.5,64)[cb]{$\mathrm{\scriptstyle t}$}
\end{picture}
\begin{picture}(170,110)(-30,0)
\Text (-5.,0)[r]{${\scriptstyle q}$}
\Text (-5.,75)[r]{${\scriptstyle \bar{q}}$}
\Text (110,0)[l]{$\mathrm{\scriptstyle \bar{b}}$}
\Text (110,15)[l]{$\mathrm{\scriptstyle \mu^-}$}
\Text (110,30)[l]{$\mathrm{\scriptstyle \bar{\nu}_\mu}$}
\Text (110,75)[l]{$\mathrm{\scriptstyle b}$}
\Text (110,45)[l]{$\mathrm{\scriptstyle \nu_e}$}
\Text (110,60)[l]{$\mathrm{\scriptstyle e^+}$}
\Vertex(88,52.5){2.0}
\ArrowLine(88,52.5)(105,45)
\ArrowLine(105,60)(88,52.5)
\Vertex(71,60){2.0}
\ArrowLine(71,60)(105,75)
\Photon(88,52.5)(71,60){2}{2}
\Text (83.8854,54.4097)[rt]{$\mathrm{\scriptstyle W^+}$}
\Vertex(88,22.5){2.0}
\ArrowLine(88,22.5)(105,15)
\ArrowLine(105,30)(88,22.5)
\Vertex(54,47.5){2.0}
\Photon(88,22.5)(54,47.5){2}{5}
\Text (70.6304,36.7774)[rt]{$\mathrm{\scriptstyle W^-}$}
\ArrowLine(54,47.5)(71,60)
\Text (60.1304,56.9726)[rb]{$\mathrm{\scriptstyle t}$}
\Vertex(37,38){2.0}
\ArrowLine(105,0)(37,38)
\ArrowLine(37,38)(54,47.5)
\Text (43.5487,46.2418)[rb]{$\mathrm{\scriptstyle b}$}
\Vertex(17,37.5){2.0}
\ArrowLine(0,0)(17,37.5)
\ArrowLine(17,37.5)(0,75)
\Gluon(37,38)(17,37.5){3}{2}
\Text (28.9,43.7488)[rb]{$\mathrm{\scriptstyle g}$}
\end{picture}
\begin{picture}(170,125)(-30,0)
\Text (-5.,0)[r]{$\mathrm{\scriptstyle g}$}
\Text (-5.,75)[r]{$\mathrm{\scriptstyle g}$}
\Text (110,0)[l]{$\mathrm{\scriptstyle \bar{b}}$}
\Text (110,15)[l]{$\mathrm{\scriptstyle \nu_e}$}
\Text (110,60)[l]{$\mathrm{\scriptstyle e^+}$}
\Text (110,30)[l]{$\mathrm{\scriptstyle \mu^-}$}
\Text (110,45)[l]{$\mathrm{\scriptstyle \bar{\nu}_\mu}$}
\Text (110,75)[l]{$\mathrm{\scriptstyle b}$}
\Vertex(84,37.5){2.0}
\ArrowLine(84,37.5)(105,30)
\ArrowLine(105,45)(84,37.5)
\Vertex(63,45){2.0}
\ArrowLine(105,60)(63,45)
\Photon(84,37.5)(63,45){2}{3}
\Text (80.,39.)[rt]{$\mathrm{\scriptstyle W^-}$}
\Vertex(42,37.5){2.0}
\ArrowLine(42,37.5)(105,15)
\ArrowLine(63,45)(42,37.5)
\Text (51.1547,45.017)[rb]{$\mathrm{\scriptstyle \bar{\nu}_e}$}
\Vertex(21,0){2.0}
\Gluon(0,0)(21,0){3}{3}
\ArrowLine(105,0)(21,0)
\Vertex(21,75){2.0}
\Gluon(0,75)(21,75){3}{3}
\ArrowLine(21,75)(105,75)
\Vertex(21,37.5){2.0}
\ArrowLine(21,0)(21,37.5)
\Text (17,18.75)[r]{$\mathrm{\scriptstyle b}$}
\ArrowLine(21,37.5)(21,75)
\Text (17,56.25)[r]{$\mathrm{\scriptstyle b}$}
\Photon(42,37.5)(21,37.5){2}{3}
\Text (31.5,33.5)[ct]{$\mathrm{\scriptstyle Z}$}
\end{picture}
\\[-1em]
\begin{picture}(170,125)(-30,0)
\Text (-5.,0)[r]{$\mathrm{\scriptstyle g}$}
\Text (-5.,75)[r]{$\mathrm{\scriptstyle g}$}
\Text (110,30)[l]{$\mathrm{\scriptstyle \bar{b}}$}
\Text (110,0)[l]{$\mathrm{\scriptstyle \mu^-}$}
\Text (110,15)[l]{$\mathrm{\scriptstyle \bar{\nu}_\mu}$}
\Text (110,45)[l]{$\mathrm{\scriptstyle b}$}
\Text (110,60)[l]{$\mathrm{\scriptstyle \nu_e}$}
\Text (110,75)[l]{$\mathrm{\scriptstyle e^+}$}
\Vertex(79,67.5){2.0}
\ArrowLine(79,67.5)(105,60)
\ArrowLine(105,75)(79,67.5)
\Vertex(53,60){2.0}
\ArrowLine(79,52.5)(105,45)
\Vertex(79,52.5){2.0}
\ArrowLine(53,60)(79,52.5)
\Text (65,52.5)[ct]{$\mathrm{\scriptstyle b}$}
\Photon(79,67.5)(53,60){2}{3}
\Text (74.5,68.)[rb]{$\mathrm{\scriptstyle W^+}$}
\Vertex(79,7.5){2.0}
\ArrowLine(79,7.5)(105,0)
\ArrowLine(105,15)(79,7.5)
\Vertex(53,15){2.0}
\ArrowLine(105,30)(79,22.5)
\Vertex(79,22.5){2.0}
\ArrowLine(79,22.5)(53,15)
\Text (65,22.5)[cb]{$\mathrm{\scriptstyle \bar{b}}$}
\Gluon(79,22.5)(79,52.5){3}{4}
\Photon(79,7.5)(53,15){2}{3}
\Text (74.5,7.)[rt]{$\mathrm{\scriptstyle W^-}$}
\Vertex(26,15){2.0}
\Gluon(0,0)(26,15){3}{4}
\ArrowLine(53,15)(26,15)
\Text (39.5,11)[ct]{$\mathrm{\scriptstyle \bar{t}}$}
\Vertex(26,60){2.0}
\ArrowLine(26,15)(26,60)
\Text (22,37.5)[r]{$\mathrm{\scriptstyle t}$}
\Gluon(0,75)(26,60){3}{4}
\ArrowLine(26,60)(53,60)
\Text (39.5,64)[cb]{$\mathrm{\scriptstyle t}$}
\end{picture}
\begin{picture}(170,125)(-30,0)
\Text (-5.,0)[r]{$\scriptstyle q$}
\Text (-5.,75)[r]{$\scriptstyle \bar{q}$}
\Text (110,0)[l]{$\mathrm{\scriptstyle b}$}
\Text (110,15)[l]{$\mathrm{\scriptstyle \nu_e}$}
\Text (110,30)[l]{$\mathrm{\scriptstyle e^+}$}
\Text (110,75)[l]{$\mathrm{\scriptstyle \bar{b}}$}
\Text (110,45)[l]{$\mathrm{\scriptstyle \mu^-}$}
\Text (110,60)[l]{$\mathrm{\scriptstyle \bar{\nu}_\mu}$}
\Vertex(88,52.5){2.0}
\ArrowLine(88,52.5)(105,45)
\ArrowLine(105,60)(88,52.5)
\Vertex(71,60){2.0}
\ArrowLine(105,75)(88.,67.5)
\Vertex(88.,67.5){2.0}
\ArrowLine(88.,67.5)(71,60)
\Photon(88,52.5)(71,60){2}{2}
\Text (84.5,53.)[rt]{$\mathrm{\scriptstyle W^-}$}
\Vertex(88,22.5){2.0}
\ArrowLine(88,22.5)(105,15)
\ArrowLine(105,30)(88,22.5)
\Vertex(54,47.5){2.0}
\Photon(88,22.5)(54,47.5){2}{5}
\Text (71.,36.)[rt]{$\mathrm{\scriptstyle W^+}$}
\ArrowLine(71,60)(54,47.5)
\Text (60.1304,56.9726)[rb]{$\mathrm{\scriptstyle \bar{t}}$}
\Vertex(37,38){2.0}
\ArrowLine(37,38)(105,0)
\ArrowLine(54,47.5)(37,38)
\Text (43.5487,46.2418)[rb]{$\mathrm{\scriptstyle \bar{b}}$}
\Vertex(17,37){2.0}
\ArrowLine(0,0)(17,37)
\ArrowLine(8.5,58)(0,75)
\Vertex(8.5,58){2.0}
\ArrowLine(17,37)(8.5,58)
\Gluon(37,38)(17,37){3}{2}
\Text (27,31)[ct]{$\mathrm{\scriptstyle g}$}
\GlueArc(57.75,-16.75)(89.5161,70.5,123.5){3}{11}
\Text (91,56)[rb]{$\mathrm{\scriptstyle \bar{b}}$}
\end{picture}
\begin{picture}(170,125)(-30,0)
\Text (-5.,0)[r]{$\mathrm{\scriptstyle g}$}
\Text (-5.,75)[r]{$\mathrm{\scriptstyle g}$}
\Text (110,0)[l]{$\mathrm{\scriptstyle \nu_e}$}
\Text (110,15)[l]{$\mathrm{\scriptstyle e^+}$}
\Text (110,30)[l]{$\mathrm{\scriptstyle b}$}
\Text (110,45)[l]{$\mathrm{\scriptstyle \bar{b}}$}
\Text (110,60)[l]{$\mathrm{\scriptstyle \mu^-}$}
\Text (110,75)[l]{$\mathrm{\scriptstyle \bar{\nu}_\mu}$}
\Vertex(78.75,67.5){2.0}
\ArrowLine(78.75,67.5)(105,60)
\ArrowLine(105,75)(78.75,67.5)
\Vertex(78.75,37.5){2.0}
\ArrowLine(78.75,37.5)(105,30)
\ArrowLine(105,45)(78.75,37.5)
\Vertex(78.75,7.5){2.0}
\ArrowLine(78.75,7.5)(105,0)
\ArrowLine(105,15)(78.75,7.5)
\Vertex(52.5,7.5){2.0}
\Gluon(0,0)(26.25,3.75){3}{4}
\Vertex(26.25,3.75){2.0}
\ArrowLine(26.25,3.75)(52.5,7.5)
\Text (39.375,1.625)[lt]{$\mathrm{\scriptstyle u}$}
\Photon(78.75,7.5)(52.5,7.5){2}{4}
\Text (68.625,3.5)[ct]{$\mathrm{\scriptstyle W^+}$}
\Vertex(52.5,67.5){2.0}
\ArrowLine(52.5,67.5)(26.25,71.25)
\Text (39.375,73.375)[lb]{$\mathrm{\scriptstyle \bar u}$}
\Vertex(26.25,71.25){2.0}
\Gluon(0,75)(26.25,71.25){3}{4}
\Photon(78.75,67.5)(52.5,67.5){2}{4}
\Text (68.625,71.5)[cb]{$\mathrm{\scriptstyle W^-}$}
\Vertex(52.5,37.5){2.0}
\ArrowLine(52.5,7.5)(52.5,37.5)
\Text (48.5,22.5)[r]{$\mathrm{\scriptstyle d}$}
\ArrowLine(52.5,37.5)(52.5,67.5)
\Text (48.5,52.5)[r]{$\mathrm{\scriptstyle d}$}
\Gluon(78.75,37.5)(52.5,37.5){3}{4}
\Text (65.625,31.5)[ct]{$\mathrm{\scriptstyle g}$}
\ArrowLine(26.25,71.25)(26.25,3.75)
\Text (22.5,37.5)[r]{$\mathrm{\scriptstyle u}$}
\end{picture}
\\[0em]
\begin{picture}(170,120)(-20,0)
\Text (-5.,0)[r]{$\mathrm{\scriptstyle g}$}
\Text (-5.,90)[r]{$\mathrm{\scriptstyle g}$}
\Text (125,0)[l]{$\mathrm{\scriptstyle b}$}
\Text (125,15)[l]{$\mathrm{\scriptstyle \nu_e}$}
\Text (125,30)[l]{$\mathrm{\scriptstyle e^+}$}
\Text (125,75)[l]{$\mathrm{\scriptstyle \bar{b}}$}
\Text (125,45)[l]{$\mathrm{\scriptstyle \mu^-}$}
\Text (125,60)[l]{$\mathrm{\scriptstyle \bar{\nu}_\mu}$}
\Text (125,90)[l]{$\mathrm{\scriptstyle g}$}
\Vertex(100,52.5){2.0}
\ArrowLine(100,52.5)(120,45)
\ArrowLine(120,60)(100,52.5)
\Vertex(100,22.5){2.0}
\ArrowLine(100,22.5)(120,15)
\ArrowLine(120,30)(100,22.5)
\Vertex(80,60){2.0}
\ArrowLine(120,75)(80,60)
\Photon(100,52.5)(80,60){2}{3}
\Text (97.5955,52.4953)[rt]{$\mathrm{\scriptstyle W^-}$}
\Vertex(80,15){2.0}
\ArrowLine(80,15)(120,0)
\Photon(100,22.5)(80,15){2}{3}
\Text (97.5955,22.4953)[rb]{$\mathrm{\scriptstyle W^+}$}
\Vertex(60,37.5){2.0}
\ArrowLine(60,37.5)(80,15)
\Text (67.0104,23.5925)[rt]{$\mathrm{\scriptstyle t}$}
\ArrowLine(80,60)(60,37.5)
\Text (67.0104,51.4075)[rb]{$\mathrm{\scriptstyle \bar{t}}$}
\Vertex(20,90){2.0}
\Gluon(0,90)(20,90){3}{2}
\Gluon(120,90)(20,90){3}{11}
\Vertex(20,37.5){2.0}
\Gluon(0,0)(20,37.5){3}{5}
\Gluon(20,90)(20,37.5){3}{6}
\Text (14,63.75)[r]{$\mathrm{\scriptstyle g}$}
\Gluon(60,37.5)(20,37.5){3}{5}
\Text (40,31.5)[ct]{$\mathrm{\scriptstyle g}$}
\end{picture}
\begin{picture}(170,120)(-20,0)
\Text (-5.,0)[r]{${\scriptstyle q}$}
\Text (-5.,90)[r]{${\scriptstyle \bar{q}}$}
\Text (125,0)[l]{$\mathrm{\scriptstyle \nu_e}$}
\Text (125,15)[l]{$\mathrm{\scriptstyle e^+}$}
\Text (125,30)[l]{$\mathrm{\scriptstyle \mu^-}$}
\Text (125,45)[l]{$\mathrm{\scriptstyle \bar{\nu}_\mu}$}
\Text (125,60)[l]{$\mathrm{\scriptstyle g}$}
\Text (125,75)[l]{$\mathrm{\scriptstyle b}$}
\Text (125,90)[l]{$\mathrm{\scriptstyle \bar{b}}$}
\Vertex(90,82.5){2.0}
\ArrowLine(90,82.5)(120,75)
\ArrowLine(120,90)(90,82.5)
\Vertex(90,37.5){2.0}
\ArrowLine(90,37.5)(120,30)
\ArrowLine(120,45)(90,37.5)
\Vertex(90,7.5){2.0}
\ArrowLine(90,7.5)(120,0)
\ArrowLine(120,15)(90,7.5)
\Vertex(60,22.5){2.0}
\Photon(90,7.5)(60,22.5){2}{4}
\Text (79.2111,10.4223)[rt]{$\mathrm{\scriptstyle W^+}$}
\Photon(90,37.5)(60,22.5){2}{4}
\Text (79.2111,32.5777)[rb]{$\mathrm{\scriptstyle W^-}$}
\Vertex(30,82.5){2.0}
\ArrowLine(30,82.5)(0,90)
\Gluon(90,82.5)(30,82.5){3}{7}
\Text (60,89.5)[cb]{$\mathrm{\scriptstyle g}$}
\Vertex(30,60){2.0}
\ArrowLine(30,60)(30,82.5)
\Text (26,71.25)[r]{${\scriptstyle q}$}
\Gluon(120,60)(30,60){3}{10}
\Vertex(30,22.5){2.0}
\ArrowLine(0,0)(30,22.5)
\ArrowLine(30,22.5)(30,60)
\Text (26,41.25)[r]{${\scriptstyle q}$}
\Photon(60,22.5)(30,22.5){2}{4}
\Text (45,18.5)[ct]{$\mathrm{\scriptstyle Z,\gamma}$}
\end{picture}
\begin{picture}(170,120)(-20,0)
\Text (-5.,0)[r]{$\mathrm{\scriptstyle g}$}
\Text (-5.,90)[r]{${\scriptstyle q}$}
\Text (125,0)[l]{$\mathrm{\scriptstyle b}$}
\Text (125,15)[l]{$\mathrm{\scriptstyle \bar{b}}$}
\Text (125,30)[l]{${\scriptstyle q}$}
\Text (125,45)[l]{$\mathrm{\scriptstyle \bar{\nu}_\mu}$}
\Text (125,90)[l]{$\mathrm{\scriptstyle \mu^-}$}
\Text (125,60)[l]{$\mathrm{\scriptstyle \nu_e}$}
\Text (125,75)[l]{$\mathrm{\scriptstyle e^+}$}
\Vertex(96,67.5){2.0}
\ArrowLine(96,67.5)(120,60)
\ArrowLine(120,75)(96,67.5)
\Vertex(72,75){2.0}
\ArrowLine(72,75)(120,90)
\Photon(96,67.5)(72,75){2}{3}
\Text (84.8069,70.5679)[rt]{$\mathrm{\scriptstyle W^+}$}
\Vertex(48,67.5){2.0}
\ArrowLine(120,45)(48,67.5)
\ArrowLine(48,67.5)(72,75)
\Text (62.8069,75.0679)[rb]{$\mathrm{\scriptstyle \nu_\mu}$}
\Vertex(96,7.5){2.0}
\ArrowLine(96,7.5)(120,0)
\ArrowLine(120,15)(96,7.5)
\Vertex(24,7.5){2.0}
\Gluon(0,0)(24,7.5){3}{3}
\Gluon(96,7.5)(24,7.5){3}{8}
\Text (60,1.5)[ct]{$\mathrm{\scriptstyle g}$}
\Vertex(24,30){2.0}
\Gluon(24,7.5)(24,30){3}{3}
\Text (18,18.75)[r]{$\mathrm{\scriptstyle g}$}
\ArrowLine(24,30)(120,30)
\Vertex(24,67.5){2.0}
\ArrowLine(24,67.5)(24,30)
\Text (20,48.75)[r]{${\scriptstyle q}$}
\ArrowLine(0,90)(24,67.5)
\Photon(48,67.5)(24,67.5){2}{3}
\Text (36,73.5)[cb]{$\mathrm{\scriptstyle Z}$}
\end{picture}
\end{center}
\caption{Typical partonic diagrams contributing to 
$\Pp\Pp\to\PW^+\PW^-\Pb\bar\Pb\to\leptfs$
at LO (first row),
to the NLO virtual corrections (second row),
and to the NLO real corrections (last row).}
\label{fi:diagrams}
\end{figure}
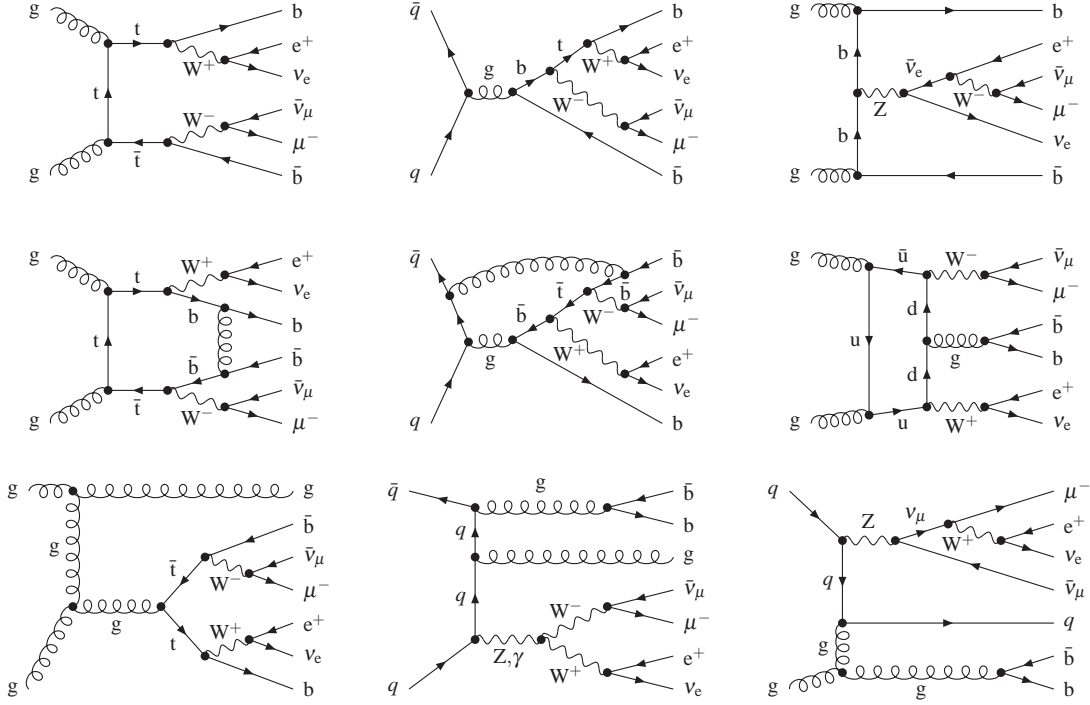

The process $\Pp\Pp\to \wwfs+X \to\leptfs+X$
describes hadronic top-quark pair production with subsequent leptonic
top-quark decays. The corresponding matrix elements
comprise doubly-resonant contributions, where the $\leptfs$
final state results from the decay of a {$\Pt\bar\Pt$} pair, as well
as singly-resonant and non-resonant diagrams, \ie contributions with
only one or no top resonance. While diagrams with intermediate
{$\Pt\bar\Pt$} states always contain two W-boson resonances from the
top-quark decays, the remaining background diagrams involve one or
two resonant W~bosons. Typical leading-order (LO) and
NLO diagrams for partonic channels with different
resonance structures are shown in \reffi{fi:diagrams}.

In the following we discuss results of our NLO QCD 
calculation~\cite{Denner:2012yc}
of the process $\Pp\Pp\to \wwfs+X \to\leptfs+X$ and of its
earlier version~\cite{Denner:2010jp}.
At LO we fully take into account all diagrams with any number of
intermediate top-quark and W-boson resonances, i.e.\ all
off-shell effects are included. For the real-emission corrections
we follow the same procedure. 
To regularize the resonances in a gauge-invariant
way we employ the complex-mass scheme~\cite{Denner:2005fg}, where the
decay widths $\Gamma_i$ ($i=\Pt,\PW$) are incorporated into the definition of
the (squared) masses, $\mu^2_i=m_i^2-\ri m_i\Gamma_i$.
All matrix elements are evaluated using the complex masses $\mu_i$, so that
resonant and non-resonant contributions are uniformly described.
In principle, it would be possible to proceed for the virtual NLO
corrections in the same way (as it was done in \citere{Bevilacqua:2010qb}),
but we decided to proceed somewhat differently.
While we still take into account all diagrams with any number of on- or off-shell
top quarks, we treat the W-boson resonances in the so-called
double-pole approximation (DPA), where only the leading contribution in an expansion
about the W~resonances is kept. This procedure reduces the set of loop diagrams
to the ones with two W~resonances only and guarantees NLO
accuracy in the vicinity of the two W-boson resonances, which includes all
regions with one or two intermediate resonant top quarks.
As discussed below, the finite-W-width (FwW) effects receive an additional suppression,
rendering this approximation rather good.
Another motivation for using the DPA resides in the possibility to include
the electroweak NLO corrections without the necessity to deal with the irreducible
$2\to6$ particle reaction; the DPA effectively splits the full process into
$\wwbb$ production of $2\to4$ type with the subsequent W~decays of $1\to2$ type.

The rest of our NLO calculation follows the standard procedure of a
diagrammatic calculation, as described in \citeres{Denner:2010jp,Denner:2012yc}
in more detail.

\section{Numerical results on finite-top-width and finite-W-width effects}

Generically, when a resonance is integrated over in an observable, finite-width
effects are of the order of $\Gamma_i/m_i$. This is nicely seen for the
finite-top-width (FtW) effects on integrated cross sections and distributions
that are dominated by the double top-quark resonance. As discussed in
\citeres{Denner:2010jp,Bevilacqua:2010qb,Denner:lh2011,Denner:2012yc} in detail for
the Tevatron and the LHC at various energies,
the FtW effects on those observables are of the typical size of ${\cal O}(\Gt/\Mt)\sim1\%$.

The situation for FwW effects is different. Observables that are dominated
by resonant $\Pt\Ptbar$ production involve W~resonances only through the
top-quark decays, and effectively any correction to these decays manifests
itself as correction to the respective top-quark branching ratio. However,
since FwW effects are universal for all top decay channels, they cancel in
the branching ratio. Thus, FwW effects come only in combination with FtW
effects in observables dominated by $\Pt\Ptbar$ production, resulting in an
additional suppression of  ${\cal O}(\Gt/\Mt\times\GW/m_{\PW})<0.5\%$.
The numerical enhancement in the number $0.5\%$, which is read off from our 
numerical study~\cite{Denner:2012yc}, results from phase-space cuts
that disturb the inclusiveness used in the power-counting argument.

In the following we exemplarily discuss FtW and FwW effects on two
distributions of phenomenological interest where those effects do not
receive the suppression described above. 
The FtW effects are quantified upon comparing the results from our
off-shell calculation with the ones obtained in the (spin-correlated)
narrow-top-width approximation (NtWA) of \citere{Melnikov:2009dn}.
The FwW effects are estimated by the difference between the two 
versions~\cite{Denner:2012yc} and \cite{Denner:2010jp} of
our calculation with off-shell and on-shell W~bosons, respectively,
as described above.
The shown results are taken from 
\citeres{Denner:lh2011,Denner:2012yc}, where all details on the
underlying input and setup 
(which is not exactly the same in \citeres{Denner:lh2011,Denner:2012yc}) 
can be found.

\begin{figure}
\includegraphics[width=\textwidth]{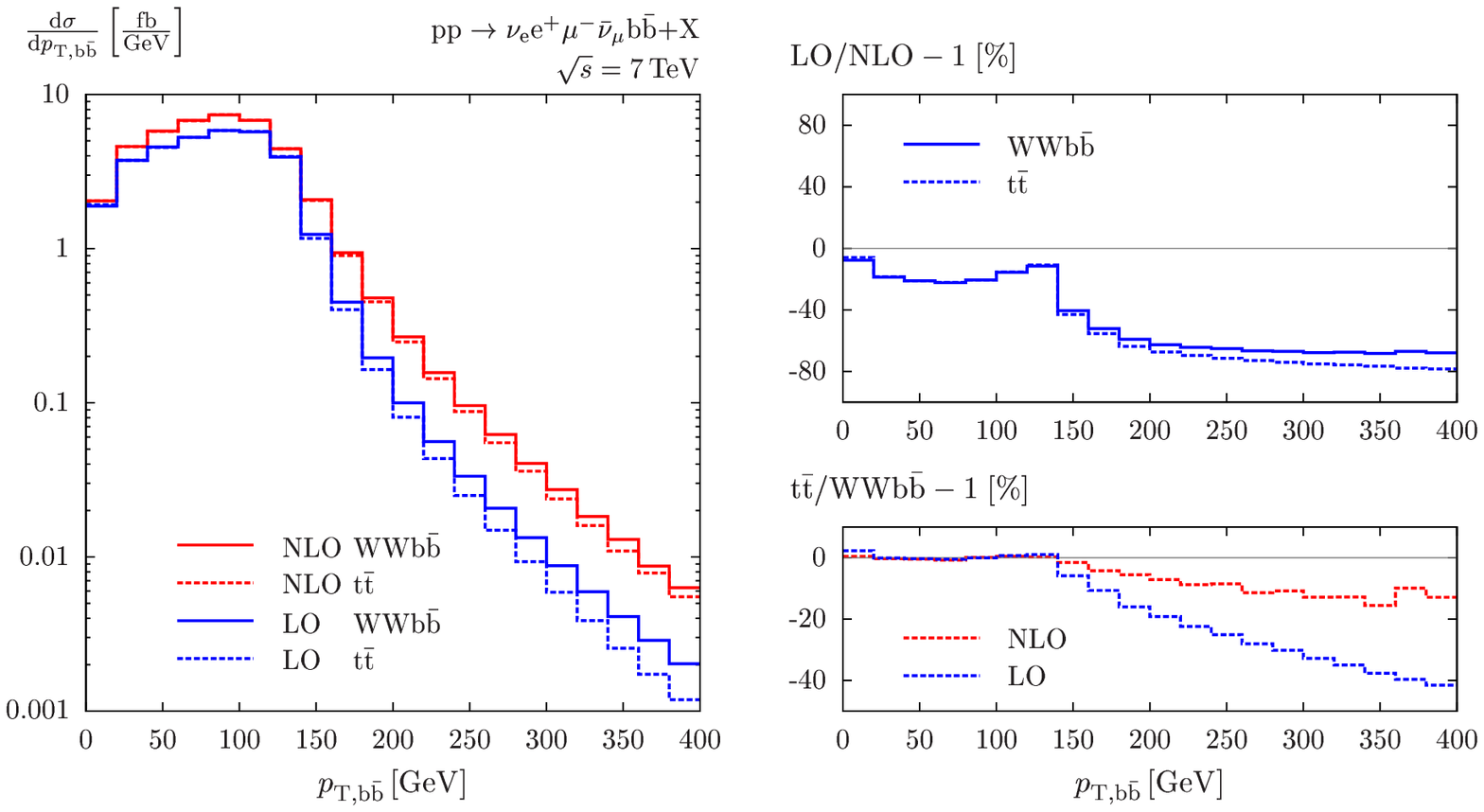}
\caption{Transverse-momentum distribution of the $\Pb\Pbbar$ pair for the LHC 
at $\sqrt{s}=7\TeV$, showing the absolute LO and NLO predictions (left),
the relative NLO corrections (upper right), and the relative difference (lower right) 
between the NtWA ($\Pt\Ptbar$) and the off-shell calculation ($\PW\PW\Pb\Pbbar$). 
(Taken from \citere{Denner:lh2011}.)}
\label{fi:ptbB_lhc7}
%
\vspace*{2.5em}
\includegraphics[width=\textwidth]{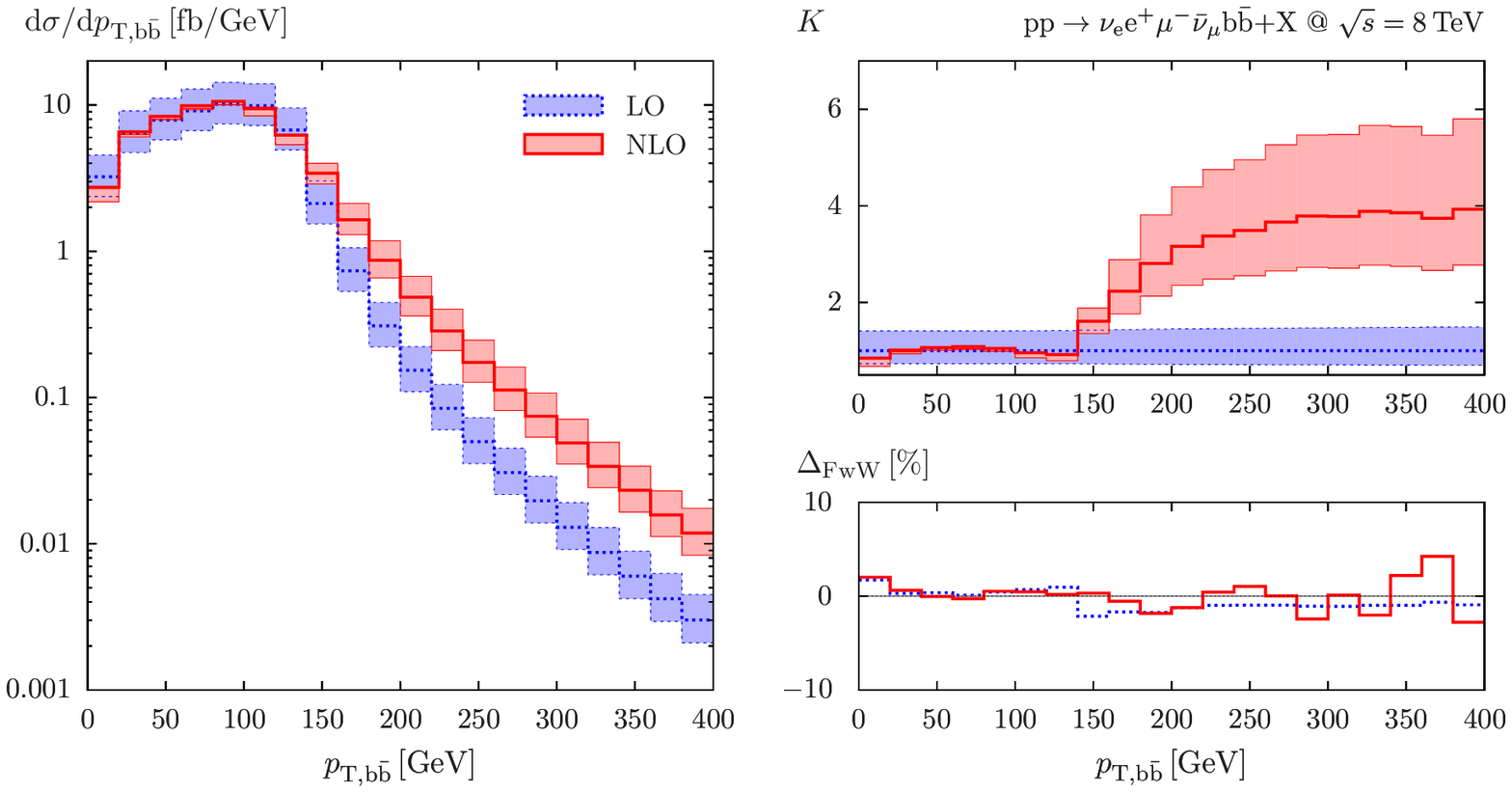}
\caption{Transverse-momentum distribution of the $\Pb\Pbbar$ pair for the LHC 
at $\sqrt{s}=8\TeV$ 
showing the absolute LO and NLO predictions with the bands illustrating the scale
uncertainty (left),
the NLO $K$ factor (upper right), and the finite-W-width effects (lower right).
(Taken from \citere{Denner:2012yc}.)}
\label{fi:ptbB_lhc8_vs}
\end{figure}

Figures~\ref{fi:ptbB_lhc7} and \ref{fi:ptbB_lhc8_vs} show the
finite-width effects on the transverse-momentum distribution of the $\Pb\Pbbar$ pair,
which plays an important role in boosted-Higgs searches with a
large $\ttb$ background. 
\begin{figure}
\includegraphics[width=\textwidth]{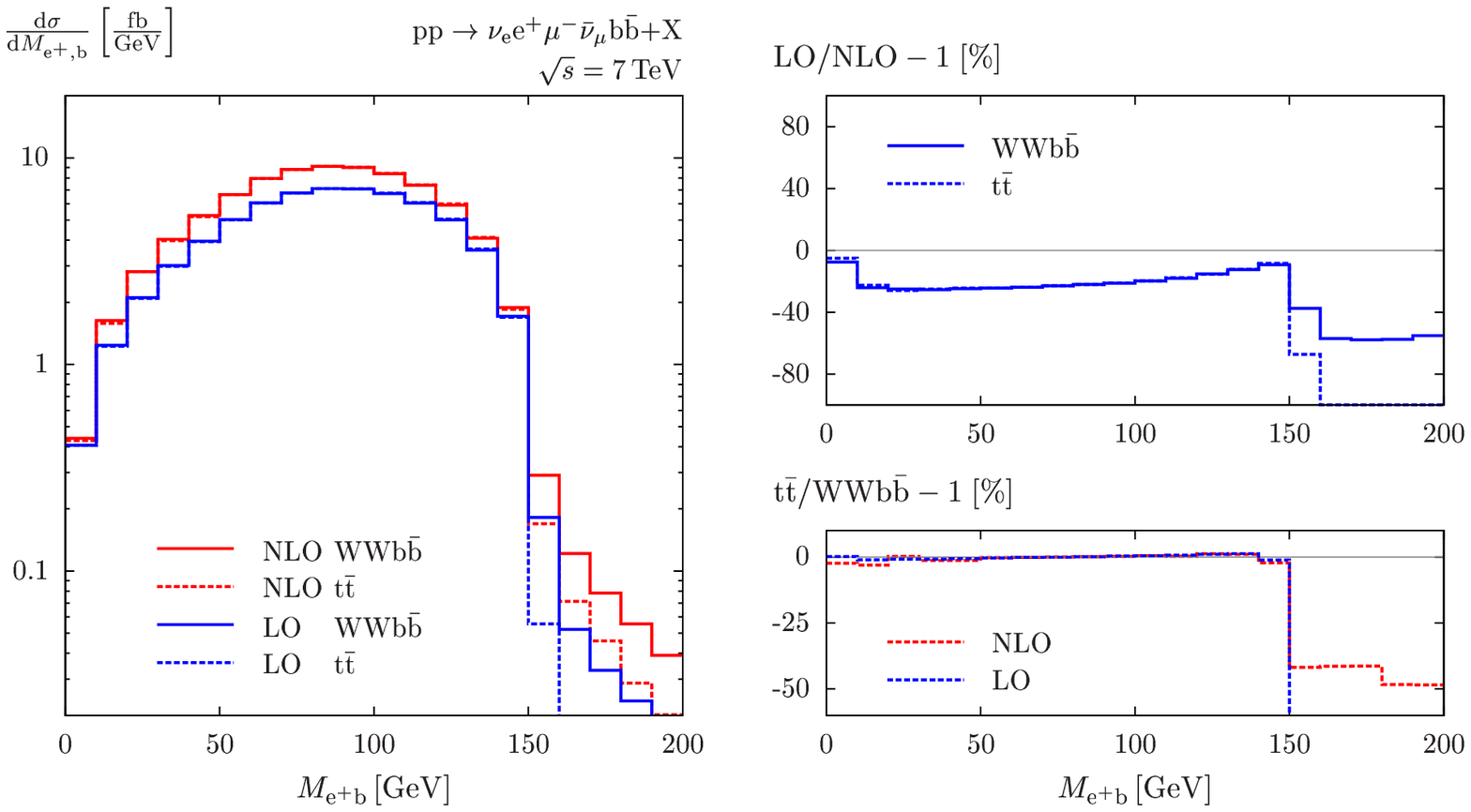}
\caption{Invariant-mass distribution of positron--\Pb-jet system for the LHC 
at $\sqrt{s}=7\TeV$, showing the absolute LO and NLO predictions (left),
the relative NLO corrections (upper right), and the relative difference (lower right) 
between the NtWA ($\Pt\Ptbar$) and the off-shell calculation ($\PW\PW\Pb\Pbbar$). 
(Taken from \citere{Denner:lh2011}.)}
\label{fi:mepb_lhc7}
%
\vspace*{2.5em}
\includegraphics[width=\textwidth]{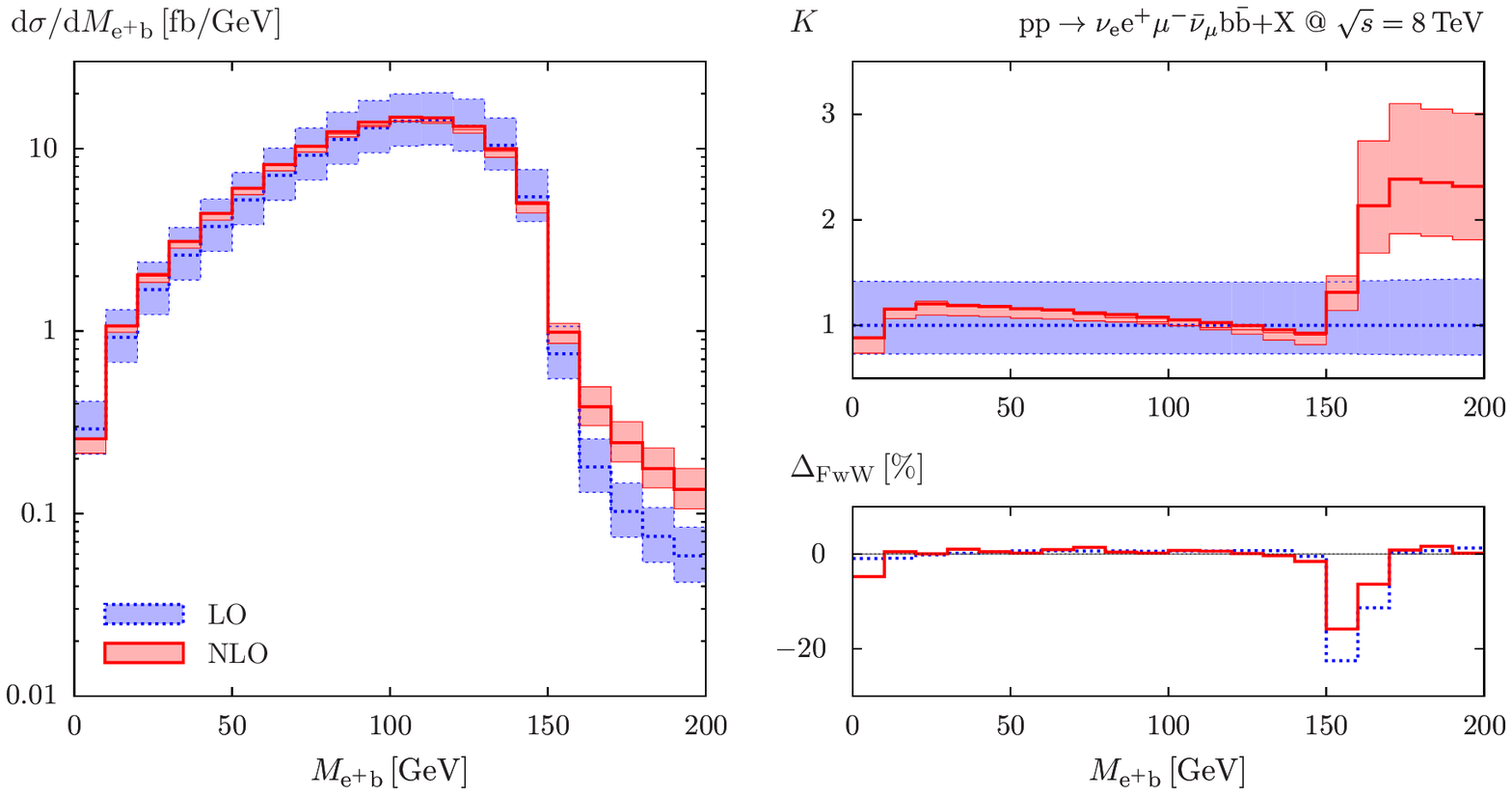}
\caption{Invariant-mass distribution of positron--\Pb-jet system 
for the LHC at $\sqrt{s}=8\TeV$ 
showing the absolute LO and NLO predictions with the bands illustrating the scale
uncertainty (left),
the NLO $K$ factor (upper right), and the finite-W-width effects (lower right).
(Taken from \citere{Denner:2012yc}.)}
\label{fi:mepb_lhc8_vs}
\end{figure}
The extraction of $\Pp\Pp\to
\PH(\to\Pb\bar\Pb)\PW$ signal at the LHC is based on the selection of
boosted $\PH\to \Pb\bar\Pb$ candidates with
$p_{\rT,\Pb\bar\Pb}>200\,\GeV$, which permits to reduce $\ttb$
contamination.
The suppression of $\ttb$
production is indeed particularly strong at $p_{\rT,\Pb\bar\Pb}\gsim
150\,\GeV$.  This is due to the fact that, for resonant $\Pt\Ptbar$ pairs, 
$\Pb$ quarks need
to be boosted via the $\pt$ of their parent (anti)top quarks in order
to acquire $p_{\rT,\Pb}> (\Mt^2-m_{\PW}^2)/(2\Mt)\simeq 65\,\GeV$, and a
$\Pb\bar\Pb$ system with high $\pt$ is kinematically strongly
disfavoured at LO, since top and antitop quarks have opposite
transverse momenta.  The NLO corrections undergo less stringent
kinematic restrictions, resulting in a significant enhancement of
$\wwbb$ events at large $p_{\rT,\Pb\bar\Pb}$.  This is clearly
reflected in the differences between the LO and NLO curves on the l.h.s.\
of the two plots.  At NLO the $\ttb$ system can
acquire large transverse momentum by recoiling against extra jet
radiation. As indicated by the upper--right plots, the NLO correction
represents $50{-}80\%$ of the cross section at high $\pt$,
corresponding to a huge $K$-factor of $2{-}4$.  
FtW effects, shown on the lower--right plot of \reffi{fi:ptbB_lhc7},
become as large as 10--30\% for
$p_{\rT,\Pb\bar\Pb}>200\,\GeV$~\cite{Denner:lh2011}. This is most
likely due to non-resonant topologies with direct
${\Pb\bar\Pb}$ production from a high-$\pt$ gluon that recoils against a
$\PW^+\PW^-$ system and splits into a $\Pb\bar\Pb$ pair.
On the other hand, FwW effects
(lower--right plot of \reffi{fi:ptbB_lhc8_vs}) stay at the level of $2\%$. 

Figures~\ref{fi:mepb_lhc7} and \ref{fi:mepb_lhc8_vs}  display
the distribution in the invariant mass of the positron
and a $\Pb$~jet, \ie the visible products of a top-quark decay.
In narrow-top-width and LO approximation this kinematic quantity is
characterized by a sharp upper bound, \eqintext{$M^2_{\Pe^+\Pb}<
  \Mt^2-m_{\PW}^2\simeq(152\,\GeV)^2$}, which renders it very sensitive to
the top-quark mass.  The value of $\Mt$ can be extracted with high
precision using, for instance, the invariant-mass distribution of a
positron and a $J/\psi$ from a B-meson
decay, an observable that is
closely related to $M_{\Pe^+\Pb}$.  In the region below the kinematic
bound, the NLO corrections to $M_{\Pe^+\Pb}$ vary between $-10\%$ and
$20\%$, and the impact of the NLO shape distortion on a precision
$\Mt$ measurement is certainly significant.  For
$M_{\Pe^+\Pb}<150\,\GeV$, the narrow-width approximation for the top quarks and W~bosons
is very good. Above the kinematic
bound, NLO corrections become clearly visible, giving rise to a tail
that extends above $M^2_{\Pe^+\Pb}=\Mt^2-m_{\PW}^2$ and also FwW
corrections become sizeable (lower--right plot of \reffi{fi:mepb_lhc8_vs}).
In this kinematic region the finite
top width causes effects at the level of 50\% (lower--right plot of 
\reffi{fi:mepb_lhc7}).
While the contribution to the total cross section from the region
above $150\GeV$ is fairly small, the impact of these contributions on
the top-mass measurement might be non-negligible, given the high $\Mt$
sensitivity of the $M^2_{\Pe^+\Pb}\simeq \Mt^2-m_{\PW}^2$ region.  A
careful comparison between NLO off-shell calculation and parton-shower
approach would be required to quantify
off-shell effects on the $\Mt$ measurement.

\providecommand{\href}[2]{#2}
\addcontentsline{toc}{section}{References}
\bibliographystyle{JHEPmod}
\bibliography{ppwwbb_ll12}

\providecommand{\href}[2]{#2}\begingroup\raggedright\begin{thebibliography}{10}

\bibitem{Schilling:2012dx}
F.-P. Schilling, {\em Int.J.Mod.Phys.} {\bf A27} (2012) 1230016,
  [\href{http://xxx.lanl.gov/abs/1206.4484}{{\tt 1206.4484}}].

\bibitem{Baernreuther:2012ws}
P.~Baernreuther, M.~Czakon, and A.~Mitov,
  \href{http://xxx.lanl.gov/abs/1204.5201}{{\tt 1204.5201}}.

\bibitem{Bernreuther:2004jv}
W.~Bernreuther, A.~Brandenburg, Z.~Si, and P.~Uwer, {\em Nucl. Phys.} {\bf
  B690} (2004) 81--137, [\href{http://xxx.lanl.gov/abs/hep-ph/0403035}{{\tt
  hep-ph/0403035}}].

\bibitem{Melnikov:2009dn}
K.~Melnikov and M.~Schulze, {\em JHEP} {\bf 08} (2009) 049,
  [\href{http://xxx.lanl.gov/abs/0907.3090}{{\tt 0907.3090}}].

\bibitem{Bernreuther:2010ny}
W.~Bernreuther and Z.-G. Si, {\em Nucl. Phys.} {\bf B837} (2010) 90,
  [\href{http://xxx.lanl.gov/abs/1003.3926}{{\tt 1003.3926}}].

\bibitem{Denner:2010jp}
A.~Denner, S.~Dittmaier, S.~Kallweit, and S.~Pozzorini, {\em Phys. Rev. Lett.}
  {\bf 106} (2011) 052001, [\href{http://xxx.lanl.gov/abs/1012.3975}{{\tt
  1012.3975}}].

\bibitem{Bevilacqua:2010qb}
G.~Bevilacqua {\em et~al.}, {\em JHEP} {\bf 1102} (2011) 083,
  [\href{http://xxx.lanl.gov/abs/1012.4230}{{\tt 1012.4230}}].

\bibitem{Denner:2012yc}
A.~Denner, S.~Dittmaier, S.~Kallweit, and S.~Pozzorini,
  \href{http://xxx.lanl.gov/abs/1207.5018}{{\tt 1207.5018}}.

\bibitem{Denner:lh2011}
A.~Denner, S.~Dittmaier, S.~Kallweit, S.~Pozzorini, and M.~Schulze, {\it
  {Finite-width effects in top-quark pair production and decay at the LHC}},
  in {\em {The SM and NLO Multileg and SM MC Working Groups: Summary Report}}
  (J.~A. Maestre, S.~Alioli, J.~Andersen, R.~Ball, A.~Buckley, {\em et~al.},
  eds.), pp.~55--63, 2012.
\newblock \href{http://xxx.lanl.gov/abs/1203.6803}{{\tt 1203.6803}}.

\bibitem{Denner:2005fg}
A.~Denner, S.~Dittmaier, M.~Roth, and L.~Wieders, {\em Nucl.Phys.} {\bf B724}
  (2005) 247--294, [\href{http://xxx.lanl.gov/abs/hep-ph/0505042}{{\tt
  hep-ph/0505042}}]. Erratum-ibid.~{\bf B854} (2012) 504--507.

\end{thebibliography}\endgroup


\end{document}